\documentclass[prd,nofootinbib,12pt]{revtex4}

\pdfoutput=1

\usepackage{color}
\usepackage[squaren]{SIunits}
\usepackage{amsmath}
\usepackage{amsfonts}
\usepackage{slashed}
\usepackage{graphicx}
\usepackage{subfigure}
\usepackage{rotating}

\usepackage{graphicx,amsmath,amssymb,epsfig,verbatim,mathrsfs,array,layout,textcomp,latexsym}
\usepackage{multirow}

\allowdisplaybreaks[1]

\renewcommand{\}}{\right\rbrace}
\renewcommand{\[}{\left\lbrack}
\renewcommand{\]}{\right\rbrack}

\newcommand{\refeq}[1]{Eq.~(\ref{eq:#1})}

\newcommand{\refsec}[1]{Section \ref{sec:#1}}

\newcommand{\bea}{\begin{eqnarray}}
\newcommand{\eea}{\end{eqnarray}}

\begin{document}

\hbox{DO-TH 14/25, QFET-2014-21}

\title{Diagnosing lepton-nonuniversality in $b \to s \ell \ell$}

\author{Gudrun Hiller}
\affiliation{Institut f\"ur Physik, TU Dortmund, Otto-Hahn-Str.4, D-44221 Dortmund, Germany}
\author{Martin Schmaltz}
\affiliation{Physics Department, Boston University, Boston, MA 02215}

\begin{abstract} 

Ratios of branching fractions of semileptonic B decays, $(B \to H \mu \mu)$ over $(B \to H ee)$ with $H=K, K^*,X_s, K_0(1430), \phi, \ldots$ are sensitive probes of lepton universality. In the Standard Model, the underlying flavor changing neutral current process $b\rightarrow s \ell \ell$ is lepton flavor universal. However models with new flavor violating physics above the weak scale can give substantial non-universal contributions. The leading contributions from such new physics can be parametrized by effective dimension six operators involving left- or right-handed quarks. 
We show that in the double ratios $R_{X_s}/R_K$, $R_{K^*}/R_K$ and $R_\phi/R_K$ the dependence on new physics coupling to left-handed quarks cancels out. Thus a measurement of any of these double ratios is a clean probe of flavor nonuniversal physics coupling to right-handed quarks. We also point out that the observables $R_{X_s}$, $R_{K^*}$, $R_{K_0(1430)}$ and $R_\phi$ depend on the same combination of Wilson coefficients and therefore satisfy simple consistency relations.

\end{abstract}

\maketitle

\section{Introduction}

Ratios of branching fractions of rare semileptonic $B$ decays into dimuons over dielectons \cite{Hiller:2003js},
\begin{align} \label{eq:RH}
R_{H} = \frac{ {\cal{B}}(\bar B \to \bar H \mu \mu) }{{\cal{B}}(\bar B \to \bar H e e) }  \, , \quad H=K, K^*, X_s,  K \pi, \ldots
\end{align}
are sensitive tests of lepton universality. The most significant theoretical and experimental uncertainties, including hadronic ones, are lepton flavor universal and drop out in the ratio, allowing for particularly clean tests of the standard model (SM).

Recently, the LHCb collaboration measured $R_K$ \cite{Aaij:2014ora}
\begin{align} \label{eq:RKdata}
R_{K}^{\rm LHCb} =0.745 \pm^{0.090}_{0.074} \pm 0.036
\end{align}
in the dilepton invariant mass squared bin $1 \, \mbox{GeV}^2 \leq q^2 < 6  \, \mbox{GeV}^2$, a deviation of $2.6 \sigma$ from universality.\footnote{Here and throughout this paper we added statistical and systematic uncertainties in quadrature.}
Previous measurements \cite{Wei:2009zv,Lees:2012tva} had significantly larger uncertainties and were consistent with the SM prediction, which is $R_K=1$ to an excellent approximation  \cite{Bobeth:2007dw}. Theory interpretations of the recent $R_K$ data are given in \cite{Das:2014sra,Alonso:2014csa,Hiller:2014yaa,Ghosh:2014awa,Biswas:2014gga,Hurth:2014vma,Glashow:2014iga,Altmannshofer:2014rta}.

Interestingly, the ratio of branching fractions in inclusive $\bar B \to X_s \ell \ell$ decays with  $q^2 > 0.04 \mbox{ GeV}^2$  \cite{iijimaLP09} (Belle) and
with  $q^2 > 0.1 \mbox{ GeV}^2$  \cite{Lees:2013nxa} (BaBar) is showing a similar trend. Both experiments find deviations from lepton universality with greater than $2 \sigma$ significance,
\begin{align} \label{eq:RXsdata}
R_{X_s}^{\rm Belle} =0.42 \pm 0.25 \, , \quad
R_{X_s}^{\rm BaBar} =0.58 \pm 0.19 \, .
\end{align}
Combining the two results is subtle for at least two reasons. One, the lower phase space boundaries in the two experiments are not exactly equal. Two, the deviations from $R=1$ seen by Belle stem from a suppression of the muon channel relative to the SM prediction whereas BaBar finds an excess of electrons relative to the SM prediction, especially in the lowest $q^2$-bin. Nonetheless, a naive error-weighted average yields
\begin{align}\label{eq:RXsave}
R_{X_s}^{\rm ave} = 0.52 \pm 0.15 \, ,
\end{align}
which is $3.1 \sigma$ away from the SM prediction%
\footnote{$R_{H}^{\rm SM}=1+{\cal{O}}(m_\mu^2/m_b^2)$ holds in general \cite{Hiller:2003js}. However,
the presence of phase space cuts induces infrared-sensitivity through collinear photon radiation at ${\cal{O}}(\frac{\alpha_e}{4 \pi} {\rm Log} (m_\ell/m_b))$ \cite{Huber:2005ig}.
Since these corrections are parametrically very small they will only be important if experimental uncertainties can be significantly improved. At present, these corrections have only been calculated for inclusive decays (implications for  $R_K$ have been commented on in \cite{Bobeth:2007dw}). Moreover,
comparison with experiment needs to be done with great care because in some analyses ${\rm Log} (m_e/m_b)$ should be replaced by ${\rm Log} (m_{\rm cut}/m_b)$, {\it i.e.} electron cut-dependent. This washes out potential differences between electrons and muons from collinear photons \cite{Huber:2008ak}.}:
$R_{X_s}^{\rm SM}=1+{\cal{O}}(m_\mu^2/m_b^2)$ \cite{Hiller:2003js}. 

While it is clearly too early to draw firm conclusions, taken at face value, (\ref{eq:RKdata}) and  (\ref{eq:RXsdata}) could be first glimpses of lepton nonuniversal physics beyond the standard model (BSM) in the flavor sector. Improved measurements as well as further crosschecks in related channels are needed to confirm and interpret these surprising results.

In this work, we draw attention to the benefits of a joint analysis of different ratios $R_H$.
Specifically, the double ratios 
\begin{align}
X_{H} \equiv \frac{R_{H}}{R_K}  \, , ~H=K^*,X_s, \phi, K_0(1430), f_0
\label{eq:doubleratios}
\end{align}
are useful high precision probes of BSM physics. They are sensitive to a different combination of couplings than $R_K$ and are therefore complementary to $R_K$.
Here, the ratios in semileptonic $\bar B_s, B_s$ decays are defined as the time-integrated untagged ratios of branching fractions
\begin{align}
R_{H} = \frac{\int_0^\infty dt {\cal{B}}(\bar B_s(t),B_s(t) \to H \mu \mu)}{\int_0^\infty dt {\cal{B}}(\bar B_s(t),B_s(t) \to H ee)}  \, , \quad  H =\phi, f_0, \eta^{(\prime)}\, , \ldots
\end{align}

The main motivation for studying several strange final states simultaneously is that the dependence on the short-distance coefficients of the $|\Delta B|=|\Delta S|=1$ effective theory is different due to the different parity of the final states. Interestingly, as we will show in the next Section, the double-ratios in \refeq{doubleratios} are only sensitive to BSM couplings to right-handed quark currents. 

Analyses of $\bar B \to \bar K^{(*)} \ell \ell$ (mostly $\mu$) and including also $\bar B_s \to \phi \mu \mu$ included right-handed quark current contributions in global fits  \cite{Descotes-Genon:2013wba,Altmannshofer:2013foa,Beaujean:2013soa,Horgan:2013pva}. The fits have some discrimination between left- and right-handed quark currents from data on angular distributions \cite{Aaij:2013qta}.
The need to cleanly disentangle QCD resonance contributions from BSM right-handed currents has recently been stressed in \cite{Lyon:2014hpa}.

The plan of the remainder of our paper is as follows: In Section \ref{sec:joint} we show that the ratios $R_H$ for different hadronic finals states provide complementary information on the $|\Delta B|=|\Delta S|=1$ couplings. We show in particular that the double ratios $X_H$ are sensitive to flavor-changing neutral currents of right-handed quarks. 
In Section \ref{sec:trans}, we show that for decays to vector mesons
$\bar B \to \bar K^* \ell \ell$ (and $\bar B_s, B_s \to \phi \ell \ell$) the transverse perpendicular $K^*$ (and $\phi$) polarizations make subdominant contributions. This observation allows us to conclude that
the double ratios $X_{K*}$ and $X_\phi$ are especially sensitive to BSM physics with couplings to right-handed quarks.
We compare $R_{K^*}$ and $R_{\phi}$ in Section \ref{sec:Rfi}, and discuss the relationship between 
$R_K$, $R_{X_s}$ and $R_{K^*}$ including CP violation in Section \ref{sec:pheno}.
We conclude in Section \ref{sec:con}. In two Appendices we give formulae and subsidiary
information.

\section{$R_H$ versus $R_K$ \label{sec:joint}}

In this Section, we give predictions for the ratios of branching fractions $R_H$ for the most promising exclusive final states which can be used to confirm and further explore potential lepton flavor nonuniversality in $b\rightarrow s\ell\ell$ decays.  We parametrize SM and BSM contributions with the usual effective Hamiltonian
\begin{align}
  \label{eq:Heff}
  {\cal{H}}_{\rm eff}= 
   - \frac{4\, G_F}{\sqrt{2}}  V_{tb}^{} V_{ts}^\ast \,\frac{\alpha_e}{4 \pi}\,
     \sum_i C_i(\mu)  {\cal{O}}_i(\mu) \, ,
\end{align}
where $\alpha_e$, $V_{ij}$ and $G_F$ denote the fine structure constant, the CKM matrix elements and Fermi's constant, respectively. We consider
 BSM physics in the following operators 
\begin{align}  \nonumber
  {\cal{O}}_{9} ^\ell& =  \bar{s} \gamma_\mu P_{L} b \, \bar{\ell} \gamma^\mu \ell \,, \quad   {\cal{O}}_{9}^{\prime \ell}  =  \bar{s} \gamma_\mu P_{R} b \, \bar{\ell} \gamma^\mu \ell \,, \\
  {\cal{O}}_{10}^\ell & = \bar{s} \gamma_\mu P_{L} b \, \bar{\ell} \gamma^\mu \gamma_5 \ell \, ,\quad  {\cal{O}}_{10}^{\prime \ell}  = \bar{s} \gamma_\mu P_{R} b \, \bar{\ell} \gamma^\mu \gamma_5 \ell \,  ,\label{eq:ops}
\end{align}
where $P_{L,R}$ are the usual chiral projectors, and we allow for lepton flavor nonuniversality with independent operators for muons and electrons, labeled here with ``$\ell$".
These operators provide the most natural explanation of $R_K<1$ \cite{Hiller:2014yaa}. Reference  \cite{Hiller:2014yaa} also showed that an alternative explanation of $R_K<1$ with (pseudo-)scalar operators involving electrons is possible but requires significant fine-tuning to avoid experimental constraints. We will not consider (pseudo-)scalar operators here.

The main point of this paper can be understood very simply from symmetries. Parity and Lorentz invariance require that the Wilson coefficients of operators with left-handed chirality $C$ (as in the SM) and their right-handed counterparts $C^\prime$ appear in the decay amplitudes of exclusive semileptonic decays in the following combinations:
\begin{align} \nonumber
C + C^\prime &: \quad K, K^*_\perp, \ldots \\
C - C^\prime &: \quad K_0(1430), K^*_{0, \parallel}, \ldots
\end{align}
Here the labels on the vector meson $K^*$ refer to its longitudinal (0), parallel $( \parallel)$ and perpendicular $(\perp)$ transversity components. One sees that contributions which only involve $C$ are universal to all decays. Therefore deviations from unity in the double ratio $X_{K^*}$ can only come from right-handed currents, $C'$.  An analogous analysis holds for
$\bar B_s, B_s$ decays.

We now give formulas for the different $R_H$. For simplicity, and to make the complementarity of the different decays manifest, the following expressions only include the dominant linear BSM contributions from interference with the SM\footnote{We also neglect contributions from the electromagnetic dipole operator which are irrelevant for the branching ratios as long as $q^2$ is not too small, and from four-quark operators which contribute at the loop level. Such contributions can and should be included once the experimental accuracy improves.  \label{foot:approx}}. Given the current experimental uncertainty of about 10\% (\ref{eq:RKdata}) these formulas are sufficiently precise. Nonetheless, for our numerical analyses and plots we use the full expressions including the quadratic BSM dependence.
\begin{align}   \nonumber 
R_K &\simeq 1 + \Delta_+  \, ,\\ \nonumber 
R_{K_0(1430)} & \simeq 1 + \Delta_-  \, ,\\ \nonumber 
R_{K^*} &\simeq 1 + p  \, (\Delta_- -\Delta_+) + \Delta_+   \, ,\label{eq:linear} \\
R_{X_s} &\simeq 1+( \Delta_+ + \Delta_- )/2\, , \nonumber \\
\Delta_{\pm} &= \frac{2}{|C_9^{\rm SM}|^2+  |C_{10}^{\rm SM}|^2} \left[ {\rm Re} \left( C_9^{\rm SM} (C_9^{\rm NP \mu} \pm C_9^{\prime \mu})^*\right) \right.\\
+ & \left. {\rm Re} \left( C_{10}^{\rm SM} (C_{10}^{\rm NP \mu} \pm C_{10}^{\prime \mu})^* \right) - (\mu \to e) \right] \,.
\nonumber
\end{align}

Notice that for inclusive decays the chirality flipped $C^\prime$ operators do not interfere with the SM, and therefore $R_{X_s}$ is only sensitive to left-chirality operators.
Note also that the presence of CP phases can suppress the SM-BSM interference. In particular, in the extreme case where the BSM Wilson coefficients are purely imaginary the interference terms vanish. We will discuss this extreme case of vanishing interference and its phenomenology in Section \ref{sec:pheno}. This requires the full expressions for $R_H$ including quadratic BSM contributions which we give in an Appendix.

Focusing again on the terms that are linear in BSM coefficients we see that the double ratios cleanly isolate right-handed currents
\begin{align} \nonumber 
X_{K_0(1430)} &  \simeq 1 + \Delta_- -\Delta_+ \, , \\
\label{eq:theX}
X_{  K^*}&  \simeq 1 + p \,  (\Delta_- -\Delta_+)  \, , \\
X_{X_s} &  \simeq 1 + \frac12 ( \Delta_- -\Delta_+ )\, . \nonumber 
\end{align}
Numerically, using SM-values at the $m_b$-scale, $C_{10}^{\rm SM} =-4.2$, $C_9^{\rm SM}=4.2$,
\begin{align} 
\Delta_- -\Delta_+ & \simeq - 0.48 \,   {\rm Re} \left(C_9^{\prime \mu} -  C_{10}^{\prime \mu}  - (\mu \to e)\right) \, .
\end{align}
These formulas show that the theoretical sensitivity to right-handed currents from scalar kaon decays, $X_{K_0(1430)}$, is two times larger than the sensitivity to right-handed currents from inclusive decays, $X_{X_s}$.
The sensitivity from vector meson $K^*, \phi$ channels depends on the ``polarization fraction" $p=p(q^2_{\rm min}, q^2_{\rm max})$, which we define as
\begin{align}
\label{eq:slopef}
p= \frac{g_{0}+g_{\parallel}}{g_{0}+g_{\parallel}+g_\perp}  \, .
\end{align}
Here $g_i=g_i(q^2_{\rm min}, q^2_{\rm max})$ denotes the $q^2$-bin dependent transversity contribution to the integrated branching ratio. Schematically,
\begin{align}   \nonumber 
{\cal{B}}( \bar B \to \bar K^* \ell \ell)& =\int_{q^2_{\rm min}}^{q^2_{\rm max}} d q^2 \frac{ d {\cal{B}}}{d q^2} \\
&= (g_0 +g_\parallel) |C -C^\prime|^2 + g_\perp |C+C^\prime|^2 \, .  \label{eq:br}
\end{align}
In the next Section, we show that $p$ is close to 1 for
the relevant $q^2$-regions, and is rather precisely known. This is good news because it implies that the vector meson channels $X_{K^*}$ are almost maximally sensitive to right-handed currents and complementary to $R_K$.

It is of course also possible to form other combinations of $R_H$ observables which are complementary to $R_K$. What is particularily nice about our double ratios $X_H$ is that deviations from unity in any of them necessarily imply new physics in right-handed currents. This is to be contrasted with the single ratios where a measured $R_{H} \neq 1$ would allow several different interpretations.

It is illuminating to rewrite our formulas in terms of Wilson coefficients  for a basis with chirally projected operators
\bea
 {\cal{O}}_{XY} ^\ell& =  \bar{s} \gamma_\mu P_{X} b \, \bar{\ell} \gamma^\mu P_{Y} \ell \quad 
{\rm with}\quad  X,Y \in \{L,R\} \ .
\eea
The $C_{XY}^\ell$ coefficients of the ${\cal{O}}_{XY} ^\ell$ are related to the standard basis as 
\begin{align}
C_{LL}^\ell & = C_9^\ell-C_{10}^\ell \, , \quad C_{LR}^\ell = C_9^\ell+C_{10}^\ell \, , \\
C_{RL}^\ell & = C_9^{\prime \ell}-C_{10}^{\prime \ell} \, , \quad C_{RR}^\ell = C_9^{\prime \ell}+C_{10}^{\prime \ell} \, .
\end{align}
In the SM, only $C_{LL}^\mu=C_{LL}^e$ are non-zero. Using the chiral basis \refeq{linear} then becomes
\bea
\Delta_{\pm} &= 2 {\rm Re}\[({C_{LL}^{ \rm NP\mu}\pm C_{RL}^\mu})/{C_{LL}^{\rm SM}}-(\mu\rightarrow e)\]\, .
\eea

It follows that while $R_K$  probes 
$ {\rm Re} \[ C_{LL}^{\mu}+C_{RL}^{\mu} -C_{LL}^{e}-C_{RL}^{e} \]$  \cite{Hiller:2014yaa},
the double ratios $X_{H}$, $H=K^*, \phi, X_s, K_0(1430),f_0$ are all only sensitive to ${\rm Re} \[ C_{RL}^\mu -C_{RL}^e\]$.
Measurement of more than one double ratio therefore over-constrains the allowed parameter space providing consistency checks. If inclusive and exclusive decays with their differing experimental and theoretical methods end up showing deviations consistent with each other the case for a short-distance interpretation with new physics will be much more convincing.

To conclude this Section we point out that since $C_9 +C_{10}=0$ in the standard model there is no sensitivity to $C_{LR}$ and $C_{RR}$ from interference with the standard model in semileptonic decays.
The latter two can be probed with leptonic decays $\bar B_s \to \ell \ell$.
We discuss correlations and constraints in Section \ref{sec:pheno}.

\section{The $K^*$ polarization fraction \label{sec:trans}}

The goal of this Section is to establish that the ``polarization fraction" $p$ defined in \refeq{slopef} is close to 1. We discuss the different kinematic regions which are sensitive to semileptonic 4-Fermi operators (\ref{eq:ops}) separately.

In the  low $q^2$-region, below the charmonium resonances, $g_0 \gg g_{\perp,\parallel}$  due to the equivalence theorem \cite{Hiller:2013cza}, hence
$p$ is of order one. Numerically, $p(1 \, \mbox{GeV}^2, 6 \, \mbox{GeV}^2)  \simeq 0.86$ \cite{Bobeth:2008ij}. Uncertainties from form factors and power corrections exist but cancel at least partially in the ratio. The precise evaluation of these uncertainties  is non-trivial as it requires knowledge of correlations between the hadronic matrix elements. Employing the form factor ratios from light cone sum rules at $q^2=0$ given in  \cite{Hambrock:2013zya} we estimate roughly ${\cal{O}}(5 \%)$ uncertainty in $p$ from form factors. This analysis can be improved by a first-principle evaluation of the matrix elements which is beyond the scope of this paper.
Note that for very low $q^2 < 1$ GeV there are contributions from resonances and the transverse polarizations  become increasingly important because of the $1/q^2$ behavior of the photon penguin contributions to the transverse amplitudes. Thus energy bins with $q^2\to 0$ are not useful for us, they are sensitive to the electromagnetic dipole operator, which is not the focus of our analysis.

For large values of $q^2$, of order $m_b^2$, above the  $\Psi^\prime$-peak, we employ the predictions from maximal $q^2$: At the point of zero hadronic recoil, $q^2_{\rm end}=(m_B-m_{K^*})^2$, end point symmetry dictates
$g_\perp=0$ and $g_0=g_\parallel/2$ \cite{Hiller:2013cza}. Hence $p=1$ exactly at zero recoil. Numerically, one finds that $p(14 \, \mbox{GeV}^2,  q^2_{\rm end} ) \simeq 0.86$. Uncertainties from form factors apply also in this kinematic region. Lattice calculations of form factor ratios applicable to the low recoil region exist \cite{Horgan:2013hoa} and suggest
similar uncertainties on $p$ as the extrapolated form factors of  \cite{Hambrock:2013zya}, about few percent.
For larger $q^2_{\rm min} \to q^2_{\rm end}$ the polarization fraction $p$ becomes  maximal, {\it i.e.}, $p \to 1$.
In intermediate $q^2$ regions the behaviour is expected to smoothly interpolate between the regions of low and maximal $q^2$.

Thus we have shown that the $K^*$ (and $\phi$) final states are dominated at low and at large $q^2$ by the ${0, \parallel}$ polarizations. The polarization fractions in the two regions are of similar size and both close to 1. As a result, decays to the vector mesons behave similarly to decays to scalars and are sensitive to right-handed currents in the combination $C-C^\prime$. This is in contrast with decays to pseudoscalars $K$ or $\eta^{(\prime)}$ where Wilson coefficients enter as  $C+C^\prime$.

Finally we mention that since the $K^*$ is analyzed as $K \pi$ 
one should consider additional resonant and non-resonant FCNC contributions to the net decay $\bar B \to \bar K \pi \ell \ell$. While  (\ref{eq:br}) remains valid in the presence of these backgrounds, they affect the accuracy to which $p$ relevant to the experimental analyses can be computed.
Contributions to all $g_i, i=0, \parallel, \perp$ can come from non-resonant decays \cite{Das:2014sra} but are phase space suppressed. Strange scalar resonances contribute to $g_0$ \cite{Becirevic:2012dp} and
are generically broad. Both contributions can be controlled with $K \pi$ mass cuts around the $K^*$ mass peak and cancel partially in the ratio $p$. In the low recoil region, we use \cite{Das:2014sra} to estimate the corrections to $p$; they do not exceed 2\%. It is also possible to remove this backgound altogether with sideband subtractions \cite{Blake:2012mb}.
The corresponding effects for $\bar B_s, B_s \to \phi (\to \bar K K) \ell \ell$  are significantly smaller  because of the $\phi$'s narrow width, and because low-lying scalar $\bar s s$ mesons either have very little overlap with the $\phi$ or have small branching ratios to $\bar K K$ \cite{Das:2014sra}.

\section{ Relations between $B_s$ and $B$ decays \label{sec:Rfi}}

We consider untagged and time-integrated branching ratios of $\bar B_s$ and $B_s$ decays to vector $\phi$, pseudoscalar $\eta^{(')}$, and scalar $f_0$ mesons. 

The formalism for $\bar B_s,B_s \to \phi \ell \ell$ was worked out in  \cite{Bobeth:2008ij}.
Using the linear BSM approximation as in \refeq{linear}, neglecting CP phases for the moment to keep the formulas manageable,
and assuming equal samples of initial $B_s, \bar B_s$ mesons, we find 
\begin{align} \nonumber
R_\phi &
\simeq  1+ \frac{(g_0+g_\parallel) (1-  y) }{(g_0+g_\parallel)(1-y)+ g_\perp (1+y)} ( \Delta_- - \Delta_+) +\Delta_+ \, , \\
& \simeq R_{K^*} - \frac{2y}{1-y} p(1-p) ( \Delta_- - \Delta_+) + {\cal{O}}((y(1-p))^2) \, ,
\label{eq:Rphi}
\end{align}
where $y=\Delta \Gamma_s/(2 \Gamma_s)$ parametrizes the impact of the lifetime difference $\Delta \Gamma_s$. Experimentally, $y=0.069 \pm 0.006$ \cite{Agashe:2014kda}.
Thus the difference between $R_{K^*}$ and  $R_\phi$ from BSM physics is an expansion in $y\,(1-p)$, with the leading correction at the \% level. In addition, we neglect very small differences between the two ratios from residual $SU(3)$-flavor breaking effects in the SM. Therefore $R_{K^*} =R_\phi$ is a very good approximation, especially in light of current experimental uncertainties on $R_K$, see \refeq{RKdata}.

Introducing CP violation through phases in the BSM Wilson coefficients changes the coefficient of the linear $y$ term in \refeq{Rphi}. The resulting formula is lengthy and we refrain from giving it here. However, the general observation that $R_{K^*}$ and $R_\phi$ are approximately equal (and the justification for it) continues to hold. Finally, note that the relation between the CP phase of the $\Delta F=2$ mixing amplitude  and the one of the $\Delta F=1$  decay amplitudes is model-dependent.
 
Turning now to $\bar B_s$ and $B_s$ decays to pseudoscalars and scalars, we find the following relationships between $B_s$ and their corresponding $B$ decay ratios:
 \begin{align}
R_{\eta^{(\prime)}} &\simeq R_K \cdot  \[ 1-2 y \sin (-\phi_M +\phi^\mu+\phi^e)  \sin (\phi^\mu-\phi^e)  \]  + {\cal{O}}(y^2) \, , \\
 R_{f_0} & \simeq  R_{K_0(1430)} \cdot  \[ 1+2 y \sin (-\phi_M +\phi^\mu+\phi^e)  \sin (\phi^\mu-\phi^e)  \]  + {\cal{O}}(y^2)  \, .
 \end{align}
 Here,  $\phi_M$ denotes the $B_s$ mixing phase and $\phi^\ell=\arg[ A(\bar B_s \to H \ell \ell)] $ for $H=\eta^{(\prime)}, f_0$.
Note that in the CP limit the ratios are equal (again ignoring residual $SU(3)$-flavor breaking).
Given that $\phi_M=0.00 \pm 0.07$ \cite{Agashe:2014kda} and $C^{\rm NP (\prime) }/C^{\rm SM} \lesssim 1/4$ the CP-induced splitting between the $R$-ratios for same-parity hadrons is below the
percent level and can be safely neglected.

\section{Scalar leptoquark UV completions \label{sec:leptoquarks}}

In this Section we briefly comment on leptoquark model building. Renormalizable models with scalar leptoquarks can easily be constructed to obtain any desired combination of Wilson coefficients $C_{XY}^\ell$ where  $\ell=e,\mu$ and $X,Y=L,R$ by choosing leptoquarks with the appropriate $SU(2)_L \times U(1)_Y$ quantum numbers. Concretely, in order to obtain the operator ${\cal{O}}_{XY}^\ell=\bar{s} \gamma_\mu P_{X} b \, \bar{\ell} \gamma^\mu P_{Y} \ell $ one chooses a leptoquark with gauge quantum numbers to allow Yukawa couplings to quarks of chirality $X$ and leptons of chirality $Y$. Schematically,
\begin{align}
\lambda_{b\ell} \,  \phi\, b_X  \ell_Y + \lambda_{s\ell} \,  \phi \, s_X  \ell_Y +{\it h.c.}
\end{align}
More precisely, the Yukawa couplings of the four possible cases $LL, LR, RL, RR$ have the fermion bilinears
$\left( ^t _b\right)_L \left(^{\nu_\ell} _\ell\right)_L,\, \overline{\left( ^t _b\right)}_L \ell_R,\,
 \overline{b}_R \left(^{\nu_\ell} _\ell\right)_L,\, b_R \ell_R$, respectively.
Integrating out $\phi$ one obtains less interesting quark flavor preserving operators and the desired operators $(b_X  \ell_Y)^\dagger s_X \ell_Y $. The latter can be rewritten as ${\cal{O}}_{XY}^\ell=\bar{s} \gamma_\mu P_{X} b \, \bar{\ell} \gamma^\mu P_{Y} \ell $ using Fierz identities. Thus for each chiral operator ${\cal{O}}_{XY}^\ell$ a corresponding scalar leptoquark UV completion can be constructed. A model with multiple Wilson coefficients turned on would require multiple different leptoquarks.

\section{Phenomenology  \label{sec:pheno}}

We discuss correlations, model-independent constraints and CP violation in Section \ref{sec:corr}, Section \ref{sec:bounds} and Section \ref{sec:CP}, respectively.

\subsection{Correlations \label{sec:corr}}

The double ratios $X_H$ receive corrections proportional to the same BSM Wilson coefficients and are therefore correlated, as can be seen clearly from \refeq{theX}.
Assuming a measurement for $X_{X_s}$ one obtains predictions for the other double ratios and vice versa,
\begin{align}  \nonumber
X_{K^*,\phi} & \simeq 1 -2 p (1- X_{X_s}) \,  , \\
\label{eq:predict}
X_{K_0(1430),f_0} & \simeq 1 -2  (1- X_{X_s}) \, . 
\end{align}
Boldly ignoring the caveats mentioned in the Introduction and combining the current data on $R_K$ (\refeq{RKdata}) and $R_{X_s}$ (\refeq{RXsave}) we obtain $X_{X_s}=0.70 \pm 0.22$. Plugging this into \refeq{predict} we predict $0.1 \lesssim X_{ K^*,\phi} \lesssim 0.9$ and 
$0.1 \lesssim R_{K^*,\phi} \lesssim 0.7$. The corresponding ranges for the scalar mesons $K_0(1430),f_0$ are very similar since $p \sim 1$ but extend to  even smaller values.%
\footnote{The formulas given here include only the BSM contributions from interference with the SM. If the experimental results on $R_K$ and $R_{X_s}$ continue to show order one deviations from unity one must include also the full BSM squared contributions. We include these in our plots.}
Note that without using the data from inclusive decays ($R_{X_s}$) one cannot currently obtain a model independent prediction for $R_{K_0(1430),f_0} $ and $R_{K^*,\phi}$ from $R_K$, they could be below or above one. However the prediction $R_{\eta^{(\prime)}} \simeq R_K$ is model independent.

We also note that for axial-vector mesons, such as the $K_1$ family, parity enforces a sign-flip in front of the Wilson coefficients for right-handed currents relative to the vectors $K^*$ or $\phi$.
Therefore, 
\begin{align} R_{K_1} &\simeq 1 + p^\prime  \, (\Delta_+ -\Delta_-) + \Delta_- \, , \\
X_{  K_1}&  \simeq 1 + (1-p^\prime) \,  (\Delta_- -\Delta_+)  \, .
\end{align}
Here we introduced a corresponding polarization factor $p^\prime$ defined as in Eq.~(\ref{eq:slopef}). While its precise numerical value will  be different from $p$ due to the difference in mass  at least, the general
expectations for $p^\prime$ are based on the same reasoning, 
hence $p^\prime \sim {\cal{O}}(1)$, and  $R_{K_1}$ is expected to be near $R_K$. Since the mixing between the nearby states $K_1(1270)$ and $K_1(1400)$ 
stems from the strong interaction it is of no concern for lepton-nonuniversality tests.

Since exclusive decays are being studied at the LHC now and high statistics data on inclusive decays will be not be collected until a few years later at Belle II, we expect that \refeq{predict} will be more effectively  used the other way around in the near future. 
That is, it will be possible to make precise predictions for $X_{X_s}$ from exclusive modes at the LHC. Especially promising are both $X_{K^*}$ and $X_\phi$ because precision studies for
$\bar B \to \bar K^* \mu \mu$ and $\bar B_s \to \phi \mu \mu$  are already available, and the results for one can serve as a crosscheck for the other.
In view of this
we discuss perspectives for a measurement of $R_{K^*}$ (and correspondingly $R_\phi$).

\begin{figure}[ht]
\begin{center}
\includegraphics[width=0.37\textwidth]{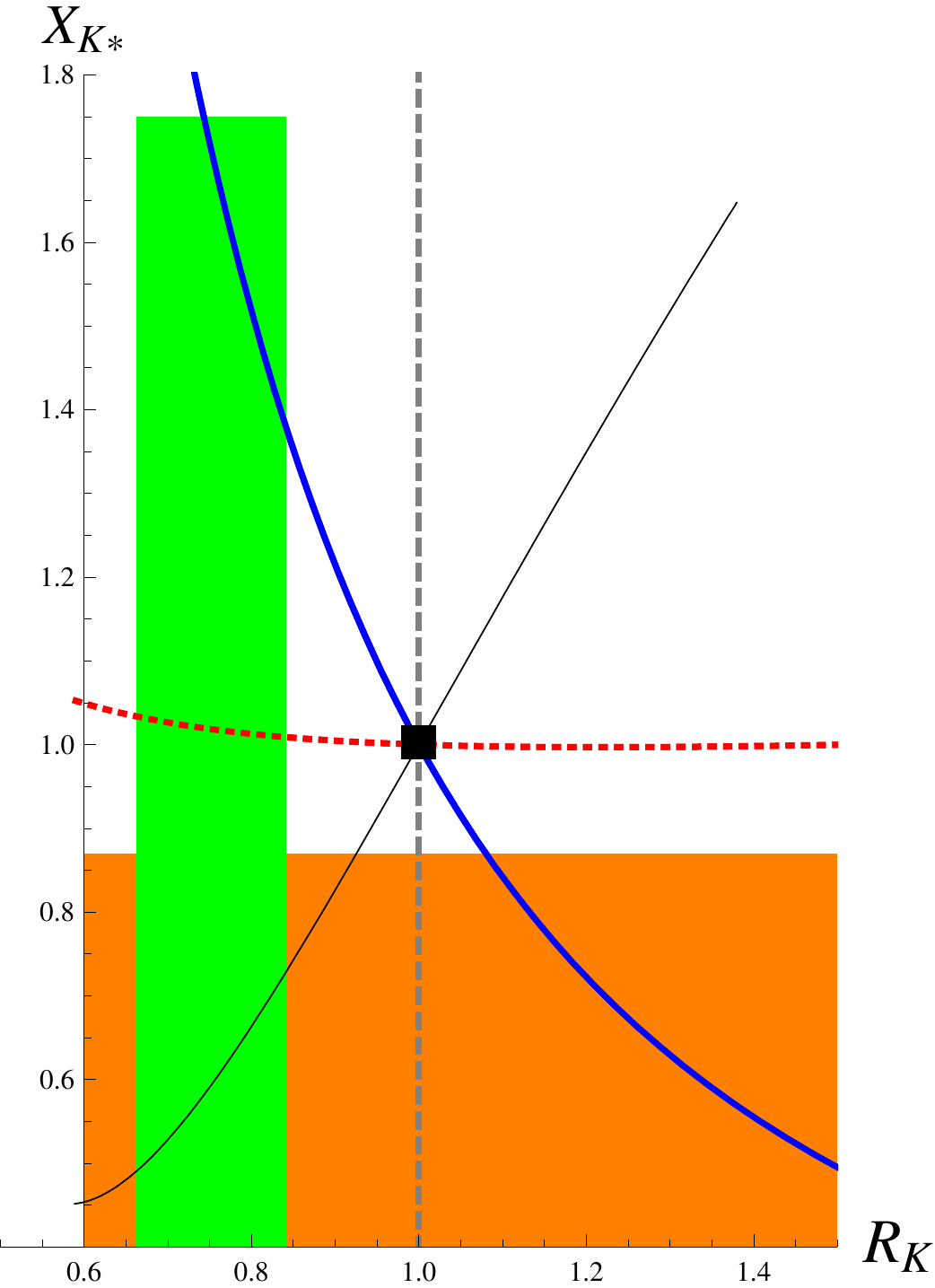}
\end{center}
\caption{\small 
$X_{K^*}$  versus $R_K$ for different BSM scenarios. The curves correspond to models in which the BSM couplings $C_{LL}^\mu$ (red dotted), $C_{RL}^e$ (blue thick solid),  $C_{RL}^e=-C_{LL}^e$  (gray dashed) and $C_{RL}^\mu=-1/2 C_{LL}^\mu$ (black thin solid) are turned on. The black square denotes lepton-universal models including the SM.
The vertical (green) shaded area denotes the $1 \sigma$ region of the $R_K$ measurement \refeq{RKdata}.
The horizontal (orange) shaded band makes use of \refeq{predict} and inclusive data \refeq{RXsdata}.
The currently allowed $1 \sigma$ region extends to even lower values of $X_{K^*} \sim 0.1$ which are not shown.}
\label{fig:corr}
\end{figure}

In Figure \ref{fig:corr} we show predictions for $X_{K^*}$  (and $X_\phi$) versus $R_K$ for $q^2=\[1,6 \] \, \mbox{GeV}^2$ in different BSM scenarios.
Plots for the large $q^2$ regions would look very similar because the polarization factors $p$ at high and low $q^2$ are very similar (\refsec{trans}).
The red dotted, blue solid and gray dashed curves correspond to predictions from different models with
BSM contributions to $C_{LL}^\mu$, to $C_{RL}^e$ and to $C_{RL}^e=-C_{LL}^e$,
respectively. 
In all scenarios considered the other BSM Wilson coefficients are set to zero. The curves in Figure \ref{fig:corr} actually hold more generally for
BSM physics which gives rise to the differences $C_{LL}^\mu-C_{LL}^e$, $C_{RL}^\mu-C_{RL}^e$, and 
$C_{RL}^\mu-C_{RL}^e=-(C_{LL}^\mu-C_{LL}^e)$ respectively. The curvature of the lines in the plot reflect the fact that we included both linear and quadratic BSM contributions to the decay rates.
All scenarios can clearly be distinguished by the two observables $R_K$ and $X_{K^*}$.

From the Figure we can see that the $R_K$ measurement taken on its own is consistent with any model with a sizeable and negative BSM contribution to ${\rm Re }(C_{LL}^\mu-C_{LL}^e)$ and all other Wilson coefficients set to zero. However, taking into account also the inclusive data, the consistency improves with an additional positive contribution to ${\rm Re}( C_{RL}^\mu-C_{RL}^e)$.  
The black thin solid curve  exemplifies such a scenario with $C_{RL}^\mu=-1/2 C_{LL}^\mu$.
In general, scenarios with  $C_{RL}^\mu=-aC_{LL}^\mu$ for $0<a<1$ are consistent with both inclusive data and $R_K$. Since $C_{LL}^\mu$ and $C_{RL}^\mu$ partially cancel each other in $R_K$, the larger the right-handed admixture $a$, the larger the required BSM Wilson coefficients.

We also observe the following correlations:
If $R_K <1 $ and  $X_{K^*} \geq 1$ (or  $R_K > 1$ and $X_{K^*} \leq 1$) then BSM dominantly contributes to $C_{LR}$. A SM-like $X_{ K^*} \simeq 1$ together with $R_K \neq 1$ points to
BSM in $C_{LL}$. If $R_K$ is SM-like but $X_{K^*} \neq 1$ then we need BSM with
$C_{LL} +C_{RL} \simeq 0$.

\subsection{Experimental constraints on the chiral Wilson coefficients\label{sec:bounds}}

In this Section we summarize the main constraints on the Wilson coefficients $C^\ell_{XY}$. In addition to the ratios $R_H$ another process which is sensitive to the Wilson coefficients is $\bar B_s \to \mu \mu$. Using the  updated combined LHCb and CMS measurement  \cite{combined}  and the SM prediction from \cite{Bobeth:2013uxa} we have
\begin{align} \label{eq:Bsmm}
\frac{ {\cal{B}}(\bar B_s \to  \mu \mu)^{exp} }{{\cal{B}}(\bar B_s \to \mu \mu)^{\rm SM} }=
0.78 \pm 0.18 \, .
\end{align}
The current 1 sigma constraints from $\bar B_s \to \mu \mu$, $R_K$ and $R_{X_s}$, respectively, read
\bea \nonumber 
0.2& \lesssim & {\rm Re}[C_{LR}^{\mu}+C_{RL}^{\mu}-C_{LL}^{\mu}-C_{RR}^{\mu}] \lesssim 1.9  \, , \\
\label{eq:constraints}
0.7& \lesssim & -{\rm  Re}[C_{LL}^{\mu}+C_{RL}^{\mu} -( \mu \to e)]  \lesssim 1.5 \, ,  \\
1.4 & \lesssim & -{\rm  Re}[C_{LL}^{\mu} -( \mu \to e)]  \lesssim 2.7  \, .\nonumber 
\eea

The available experimental and theoretical analyses on $\bar B \to \bar K^{(*)} \mu \mu$ are much more sophisticated than their electron counterparts, and are subject to correlations \cite{Descotes-Genon:2013wba, Altmannshofer:2013foa, Beaujean:2013soa,Horgan:2013pva}, and more recently 
 \cite{Ghosh:2014awa,Hurth:2014vma,Altmannshofer:2014rta}.
Adopting the results of \cite{Hurth:2014vma} which include $R_K$ data but assume real Wilson coefficients, the allowed ranges at 95 \% CL are roughly
\begin{align}
-2.7 \lesssim & C_{LL}^{\mu} \lesssim 0 \, , \quad \quad 
-2.0 \lesssim C_{RL}^{\mu} \lesssim 0.6 \, , \\
-16.8 \lesssim & C_{LL}^{e} \lesssim 2.5 \, , \quad \quad 
-10.9 \lesssim C_{RL}^{e} \lesssim 10.9  \, . 
\end{align}
LHCb's upcoming $3 \mbox{fb}^{-1}$ analysis of $\bar B \to \bar K^{*} \mu \mu$ data should significantly improve the muon bounds.

\subsection{CP violation \label{sec:CP}}

In general, one would not expect the phases of BSM physics to be aligned with the CKM phases. It is clear from inspecting \refeq{linear} that CP phases in BSM Wilson coefficients suppress the impact of new physics on the ratios $R_{H}$. On the flip side, interesting CP violating observables become measurable. In this Section, we highlight some of these observables and discuss the impact of CP phases on $R_K$ and $X_H$.

To begin, we generalize \refeq{linear} to include the quadratic BSM contributions in the chiral basis
\begin{align}
\label{eq:quad}
R_K & \simeq  \left(1 + 2\,  {\rm Re} \left[ \frac{C_{LL}^\mu+C_{RL}^\mu}{C_{LL}^{SM}}\right]
+ \frac{|C_{LL}^ {\rm NP \mu }+ C_{RL}^{\mu }|^2  +|C_{LR}^{ \mu }+C_{RR}^{\mu} |^2}{|C_{LL}^{\rm SM}|^2}\right)/
\left(\mu \to e\right) \, ,
\end{align}
and similar expressions for the other $R_H$, see Appendix A.
One sees that when $C^{\rm NP}\ll C^{\rm SM}$ this is generically dominated by the linear BSM terms which we discussed in the previous Sections. Thus our previous discussions of $R_H$ and $X_H$ also apply to the case of large CP violation.

However, an interesting exception arises when the BSM Wilson coefficients $C_{LL}^\ell$ and $C_{RL}^\ell$ are close to pure imaginary so that the quadratic terms in \refeq{quad} dominate.  In that case we obtain for the double ratios
\begin{align}
X_{H}-1 & \propto
 -4  \,  {\rm Re} \left( C_{LR}^ { \mu}  C_{RR}^{\mu * } +C_{LL}^ {\rm NP \mu}  C_{RL}^{ \mu * } -(\mu \to e)\right)/|C_{LL}^{\rm SM}|^2 \, 
\end{align} 
which is non-zero only if there are new physics contributions to both left- and right-handed quark currents. When BSM only enters quadratically, one would normally expect the deviations of $X_H$ or $R_H$ from one to be small. However, current data still allow that the large deviations from the SM in \refeq{RKdata} are due to quadratic BSM physics. Since the quadratic contributions from NP in muons give the wrong sign for $R_K$, this would require large pure imaginary BSM Wilson coefficients for electrons. Comparing to the 1 sigma range in \refeq{RKdata} we find
 \begin{align}
11.2 \lesssim |C_{LL}^ {\rm NP e }+ C_{RL}^{e }|^2  +|C_{RR}^{e }+ C_{LR}^{e} |^2  \lesssim   23.8 \,,
\end{align}
requiring almost order one BSM contributions relative to the SM, $C^{\rm NP} \sim 1-4$.
This is allowed at 2 sigma, but is excluded at 1 sigma by data on the $\bar B \to \bar Kee$ branching ratio at low $q^2$ \cite{Aaij:2014ora,Bobeth:2012vn}.

In the following, we give predictions for CP sensitive observables which are useful for diagnosing lepton nonuniversal CP violation.
%
CP-phases can be probed with CP-asymmetries in $\bar B \to \bar K^* \ell \ell$ and $\bar B_s, \bar B_s \to \phi \ell \ell$  \cite{Bobeth:2008ij}.
These asymmetries are effectively nulltests of the SM because of  the smallness of
$V_{ub} V_{us}^*/(V_{tb} V_{ts}^*)$.
The asymmetries $A_{7,8,9}^{(D)}$, which can be obtained from an angular analysis (see Appendix \ref{sec:aa}) are particularly promising because they are naive T-odd and hence do not require strong phases. 
They exhibit the following features: $A_9$ is sensitive to right-handed currents, $A_7^{D}$ is sensitive to $C_{10}^{(\prime)}$, and $A_{7,8}^{(D)}$ are CP-odd and can be obtained without flavor-tagging. The latter is advantageous for $B_s$ decays which are not self-tagging.

\begin{figure}[ht]
\begin{center}
\includegraphics[width=0.47\textwidth]{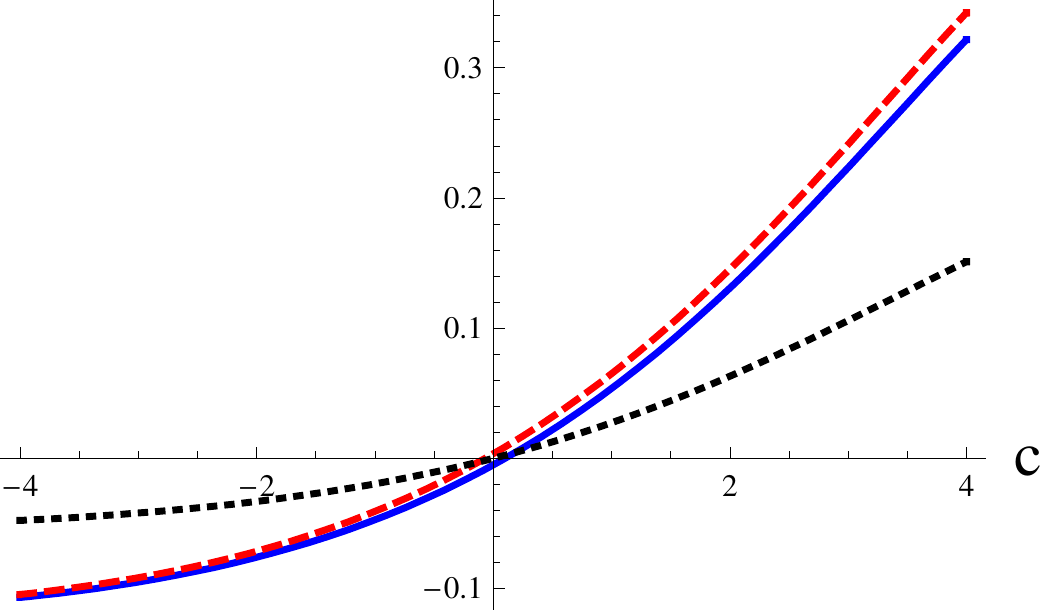}
\includegraphics[width=0.47\textwidth]{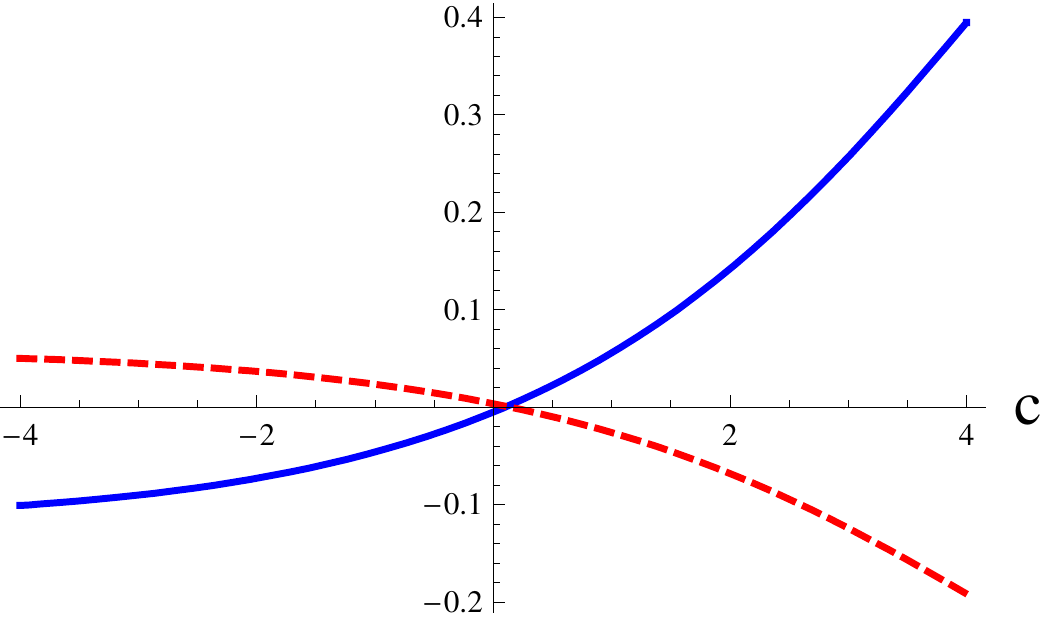}
\includegraphics[width=0.47\textwidth]{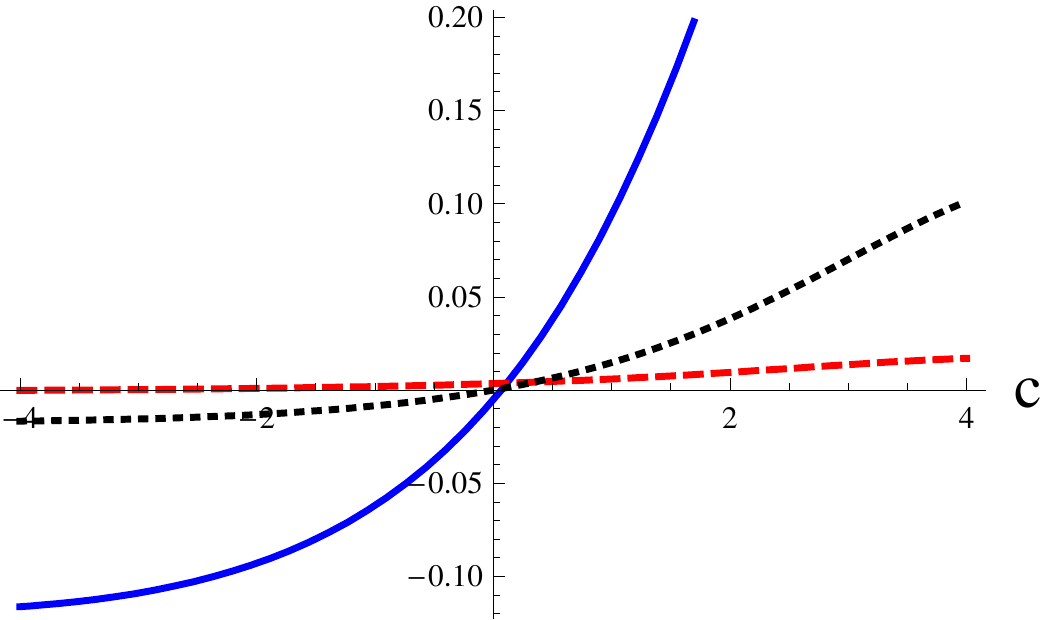}
\end{center}
\caption{\small 
T-odd CP asymmetries in $\bar B \to \bar K^* \ell \ell$  decays for $q^2=\[1,6 \] \, \mbox{GeV}^2$.
Blue solid, red dashed, black dotted curves correspond to $A_{7,8,9}^{(D)}$, respectively. The upper plot to the left is 
showing $A_{7,8,9}^{(D)}$ as a function of the parameter $c$ in the BSM Wilson coefficient $C_{RL}^e = - c\, e^{i \pi/3}$.
The plot on the upper right is for BSM in $C_{LL}^e = c\, e^{i \pi/3}$.  In this case - as in the SM -
$A_9$ is negligible because of the absence of right-handed currents. 
The third (lower) plot interpolates between these limiting cases with  $C_{LL}^e=-2 C_{RL}^e = c\, e^{i \pi/3}$.
Plots for other values of the CP phase show similar behavior with asymmetries scaling roughly as the sine of the CP-phase.
These plots apply to  $\bar B \to \bar K^* \mu \mu$ as well, however the magnitudes of the Wilson coefficients for muons are more strongly constrained than in $\bar B \to \bar K^* ee$.}
\label{fig:ACP}
\end{figure}

In Figure  \ref{fig:ACP}  we show  $A_{7,8,9}^{(D)}$ for $q^2=\[1,6 \] \, \mbox{GeV}^2$ in three scenarios with BSM physics coupling to electrons.
The CP asymmetries can be sizable, reaching ${\cal{O}}(\mbox{few} \, 10 \%)$.
By comparing the asymmetries in the plots on the upper left  ($C_{RL}^e$) , upper right  ($C_{LL}^e$) and below
($C_{RL}^e=-C_{LL}^e/2$) the scenarios can clearly be distinguished.
In particular, an opposite sign between $A_{7,8}^{(D)}$ together with vanishing $A_9$ would point  to BSM in $C_{LL}^e$ only.
The parametric behaviour of the CP asymmetries also holds for $\bar B \to \bar K^* \mu \mu$. However, one must keep in mind that the allowed ranges for the magnitudes of the Wilson coefficients are much more constrained in the muon case.

Note that in the approximations stated  in Footnote \ref{foot:approx} the asymmetry $A_7^{D}$ vanishes. It arises from interference between $C_{10}-C_{10}^\prime$ and the electromagnetic dipole operator, which we assumed in our numerical analysis to be SM-valued. We emphazise that in all plots we take into account the full expressions including all non-vanishing SM Wilson coefficients as well as linear and quadratic BSM physics.
Finally, we remark that $A_7^{D}$ vanishes at high $q^2$ because of features of the lowest order OPE \cite{Bobeth:2012vn}. This is true for arbitrary Wilson coefficients for any of the operators in \refeq{ops}.

\section{Conclusions \label{sec:con}}

We proposed a combined study of the ratios $R_{H}$ for final state hadrons $H=K,K^*,X_s,\phi,K_0(1430),f_0,\eta^{(')}, K_1, ...$ to help decipher BSM effects in $b \to s \ell \ell$ transitions. While each of the ratios $R_H$ is sensitive to lepton-nonuniversality, the $R_H$ depend on only two combinations of short-distance Wilson coefficients, $\triangle_+ +\Sigma_+$ and $\triangle_- +\Sigma_-$. Since there are more than two possible final states which are experimentally accessible this allows one to test for consitency between the different measurements. In particular, we have pointed out the importance of this cross check between inclusive and exclusive decays. Since these employ very different experimental and theoretical methods, global consistency between them could make a convincing case for BSM physics. 
To leading order in new physics, $\Sigma_\pm$ are negligible, and the $\triangle_\pm$ simplify to 
\bea
\triangle_+ +\triangle_-\simeq 0.48\; {\rm Re}\,(C_{LL}^\mu-C_{LL}^e) \quad\quad {\rm and}\quad\quad  \triangle_+ -\triangle_-\simeq 0.48\; {\rm Re}\,(C_{RL}^\mu-C_{RL}^e)\ .
\eea

We further showed that the double ratios $X_H\equiv R_H/R_K$ 
depend only on lepton nonuniversality in right-handed currents, $C_{RL}^\mu-C_{RL}^e$.
We point out that the polarization fraction of the decays $\bar B \to \bar K^{*} \ell \ell$ and $\bar B_s \to  \phi \ell \ell$ makes them particularly sensitivity to BSM physics and complementary to $R_K$,  as shown in Figure \ref{fig:corr}.

Current data suggest $R_K <1$ (\refeq{RKdata})  and  $R_{X_s} <1$ (\refeq{RXsdata}), pointing to BSM-physics in the SM-like chirality operator
${\rm Re }(C_{LL}^\mu-C_{LL}^e)<0$ and possibly also some right-handed contributions ${\rm Re}( C_{RL}^\mu-C_{RL}^e)>0$. Interestingly, the currently preferred region of parameter space has BSM Wilson coefficients which are not much smaller than the SM ones. 

Taking the current data on $R_K$ and $R_{X_s}$ at face value we predict $0.1 \lesssim R_{K^*,\phi} \lesssim 0.7$ from \refeq{predict} which reflects that double ratios $X_H$ all depend on the same BSM parameter. This prediction should clearly be taken with a grain of salt, possibly indicating a trend, rather than as a precision prediction, since \refeq{RKdata} and \refeq{RXsdata} are based on respective frontier measurements.
While the prediction is currently still hampered by large experimental uncertainties, it is a nice example of what can be done with this program of measurements.
Future data together with  dedicated fits of the global $|\Delta B|=|\Delta S|=1$ will be able to clarify if these preliminary patterns are real.
We note that without taking into account the inclusive data on lepton-nonuniversality, (\refeq{RXsdata}), $R_H-1$ for $H=K^*,\phi, K_0(1430),f_0$ could still have either sign.

At present, experimental analyses from the LHC of $B$ and $B_s$ decays to final states with muons are much more advanced than the corresponding analyses for electrons. Hence much larger BSM effects are still allowed in the Wilson coefficients of electrons. To improve on the existing searches for lepton nonuniversality, improved studies of decays to electrons are especially important.

\acknowledgments
GH is happy to thank Tom Blake, Christoph Bobeth, Tim Gershon and Kostas Petridis for useful communications.
This work is supported in part  by the DFG Research Unit FOR 1873 ``Quark Flavour
Physics and Effective Field Theories" (GH) and by the US Department of Energy Office of Science under Award Number DE-SC-0010025 (MS). \\

\appendix
\section{Ratios and double ratios}

The ratios $R_H$ and double ratios $X_H$ have BSM contributions from interference with the SM ($\Delta_\pm$), and from pure BSM-squared terms ($\Sigma_\pm$):
\begin{align}  \label{eq:quada}
R_K &\simeq 1 + \Delta_+  + \Sigma_+  \, , \\   
R_{K_0(1430)} &\simeq 1 + \Delta_-  + \Sigma_-  \, , \\
 \label{eq:RKstquad}
R_{K^*} &\simeq1+  p  \, (\Delta_- -\Delta_+ + \Sigma_- -\Sigma_+ ) + \Delta_+ + \Sigma_+  \, , \\  
R_{X_s} &\simeq1+    (\Delta_- -\Delta_+ + \Sigma_- -\Sigma_+ )/2 \, , \\  
X_{K_0(1430)}& \simeq1+ (\Delta_- -\Delta_+ + \Sigma_- -\Sigma_+ ) \, , \\  
X_{K^*}& \simeq1+ p \,  (\Delta_- -\Delta_+ + \Sigma_- -\Sigma_+ ) \, , \\  
X_{X_s}& \simeq1+ (\Delta_- -\Delta_+ + \Sigma_- -\Sigma_+ )/2 \, ,   \label{eq:quadb}
\end{align}
where $\triangle_\pm$ were defined in \refeq{linear} and 
\begin{align}
\Sigma_{\pm}= \frac{|C_9^ {\rm NP \mu } \pm  C_9^{\prime \mu }|^2  +|C_{10}^{\rm NP \mu } \pm C_{10}^{\prime \mu} |^2}{|C_9^{\rm SM}|^2+  |C_{10}^{\rm SM}|^2}  - (\mu \to e) \, .
\end{align}

\section{$\bar B \to \bar K^*(\to \bar K \pi)  \ell \ell$ angular distribution \label{sec:aa}}

The CP asymmetries $A_{i}^{(D)}$ related to the coefficients $J_i$ and $\bar J_i$ of the $\bar B \to \bar K^* \ell \ell$  angular distribution 
of $\bar B$ and CP conjugate decays
\begin{align}
\frac{d^4 \Gamma(\bar B \to \bar K^* (\to \bar K \pi)\ell \ell) }{d q^2 d^3  {\rm angles}} = \sum_{i=1s,1c,2s,2c,3,..9} J_i(q^2) f_i({\rm angles}) \, , \\
\frac{d^4 \bar \Gamma( B \to K^*(\to K \pi)  \ell \ell) }{d q^2 d^3  {\rm angles}} = \sum_{i=1s,1c,2s,2c,3,..9} \bar J_i(q^2) f_i({\rm angles}) \, ,
\end{align}
 are defined as
  \begin{align}
 A_{i}^{D} & =-2 \frac{\int d q^2 (J_i-\bar J_i) }{ \int d q^2 (d \Gamma/dq^2 + d \bar \Gamma/dq^2)}  \,, ~~ i=4,5,7,8  \, ,\\
  A_{j}  & =2 \frac{\int d q^2 (J_j-\bar J_j) }{ \int d q^2 (d \Gamma/dq^2 + d \bar \Gamma/dq^2)}  \,, ~~ j=3,6,9 \, .
 \end{align}
See \cite{Bobeth:2008ij} for details including the definition of the three angles and the known trigonometric functions $f_i$. Analogous definitions hold for $\bar B_s,B_s \to \phi \ell \ell$ decays. Note that $d \Gamma / dq^2=2 J_{1s}+J_{1c}-(2 J_{2s} +J_{2c})/3$.
 {}From  the composition of the $J_i$ in terms of transversity amplitudes  we can extract the dependence on right-handed currents. Schematically,
 within the approximations stated in Footnote \ref{foot:approx}, 
 \begin{align}
 J_3 & \propto   |C+C^\prime|^2 -|C-C^\prime|^2 =4 {\rm Re} (C C^{\prime *}) \, , \\
 J_4 & \propto  {\rm Re} \left( (C-C^\prime) (C-C^\prime)^*\right) =|C- C^\prime|^2 \, , \\
 J_{5,6} & \propto  {\rm Re} \left( (C-C^\prime) (C+C^\prime)^*\right) =|C|^2-|C^\prime|^2 \, , \\
 J_7 &\propto  {\rm Im} \left( (C-C^\prime) (C-C^\prime)^*\right) =0 \, , \\
 J_{8,9} & \propto  {\rm Im} \left( (C-C^\prime) (C+C^\prime)^*\right) =2 {\rm Im} (C C^{\prime *}) \, .
 \end{align}
 This shows that it is possible to construct simple CP-(a)symmetric observables for diagnosing lepton-nonuniversality along the lines reported in this paper.
 Note however, that neglecting dipole operators is less justified for the $J_{3,..., 9}$ at lower $q^2$ than it is for the branching ratio. This is because the branching ratio has a large contribution from longitudinal $K^*$ in which the dipole contribution is not enhanced by $1/q^2$. On the other hand, the dipole operator in both transverse amplitudes is $1/q^2$ enhanced and causes the famous zero in the forward-backward asymmetry $\propto J_6$, {\it i.e.} the  vector-coupling to leptons, around $q^2 \sim (3-4) \, \mbox{GeV}^2$.
 
 Focussing on the interesting  T-odd CP asymmetries,
 neglecting the dipole contributions is justified at large $q^2$ and predicts that $A_7^D$ vanishes while $A_8^D$ and $A_9$ have identical dependences on Wilson coefficients. 
 Both predictions are modified by contributions from the
 electromagnetic dipole operator to $J_7$ and $J_8$, while $A_9$ remains zero iff $C^\prime=0$.

\end{document}